# New Transition Mechanism: The Shaking Effect

C. Montero, J. García Ureña and M. Dorado [a]

It has been observed that the trajectory of a $H_2$ molecular beam is modified as it passes through a homogeneous magnetic field and an oscillating magnetic field. Based on this observation, this work describes a new method to determine the magnetic moment of atoms and molecules with internal angular momentum. This new method is similar to that used in the Molecular Beam Magnetic Resonance (MBMR) but 'it does not employ' the inhomogeneous fields *A* and *B*. Results shown here are in agreement with the theory of the motion of systems with internal angular momentum developed by M.Dorado. Furthermore, in this work the transition mechanism given by I.I. Rabi and N.F. Ramsey is reviewed and completed with the introduction of nutation frequency and shaking effect.

## 1. Introduction

In 1921 O. Stern and W. Gerlach [1-2] experimentally demonstrated the existence of the spatial quantization of the angular momentum proposed by A. Sommerfeld. As a consequence, the angular momentum of an atom with a given electronic configuration could be determined. Importantly, this finding allowed the development of techniques such as the MBMR, which is one of the mayor contributions to the experimental study of the atomic and molecular structure [3-11].

The MBMR method (developed by I.I. Rabi) measures, in the presence of a magnetic field, the Larmor frequency of an atom or molecule. This is achieved by forcing a molecular beam to pass through two consecutive inhomogeneous magnetic fields, *A* and *B*, which have opposite directions. Thus the molecular beam describes a sigmoidal trajectory and is then collected by a detector (see Figure 1) [12]. Between these two fields, there is also an homogeneous magnetic field, *C*, which induces the spatial quantization of the magnetic momentum of the atom or molecule under study. Additionally, in this same region, *C*, an oscillating magnetic field is applied which induces transitions when its frequency is similar to the Larmor frequency.

MBMR is based on two main principles: Firstly, the trajectory deflection of an atom with magnetic moment in the presence of a inhomogeneous magnetic field. Secondly, the non-adiabatic transitions produced on a system with angular momentum when a oscillating field in region *C* has a frequency similar to the Larmor frequency.

Rabi explained this effect in terms of spatial 'reorientation' of the angular momentum as the transition occurs. He proposed that, due to the interaction with the oscillating field, a particle in state *m* makes a transition to state *m'* and therefore, in the inhomogeneous field *B*, it describes a trajectory different to that it would have followed in state m, hence it does not reach the detector.

The main aim of this work is to show that this 'reorientation' phenomenon may be explained in a different manner which is in good agreement with the experimental and theoretical results reported by Rabi and others [13 - 17]. In addition, the equations that describe the particle trajectory during the resonant process are shown and preliminary results are discussed. This new deflection mechanism predicted by M.Dorado's theory has been reported recently for $NO_2$ [18], NO [19-22], NO dimmer [23], Ba..$FCH_3$ cluster [24], and Tolueno[25] in MBER experiments without *A*- and *B*- fields.

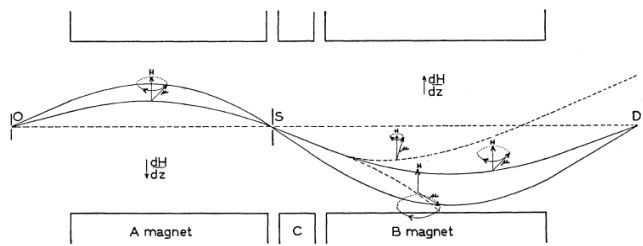

**Fig. 1.** Typical path of molecules in a M.B.M.R. experiment. The two solid curves show the paths of two molecules having different moments and velocities and whose moments do not changed when passing through the magnets. This is indicated by the small gyroscopes drawn on one of these curves, in which the projection of magnetic moment along the field remains fixed. The two dotted curves in the region of the *B* magnet show the paths of two molecules whose projection of the nuclear magnetic moments along the field changes in the region of the *C* magnet. (From reference 3).

[a] *Rotation & Torque Research Center. Scientific Pool of Madrid. Pol. Ind. Tres Cantos Oeste s/n .C.P: 28760. Tres Cantos, Madrid (Spain). Fax: +0034 918 049 913   Tel: +0034 918 049 912  Mail: cmontero@cirta.es*

## 2. Theoretical Transition Mechanism

A magnet situated in *C* generates a homogeneous magnetic field, *H₀*. Also in this region, there is a magnetic field *H₁* of lower intensity and perpendicular to *H₀*. From a classic point of view, an atom with magnetic moment, *μ*, and angular momentum, *L*, in the presence of a magnetic field *H₀* feels a torque that induces the precession of the magnetic moment along the direction of the applied field and this precession frequency is known as the Larmor frequency

$$\omega_{Larmor} = \frac{\mu H_0}{L \hbar} \qquad (1)$$

with *L* in ℏ units. From the above expression it can be deduced that, if both the values of *H₀* and the Larmor frequency are known it is easy to calculate *μ*. The result of this formalism may be summarised as follows. If the frequency of the oscillating field **H₁** is different from the Larmor frequency the atom will remain quantized, with respect to the direction of the field **H₀**, in its initial state *m*, characterized by a total angular momentum *F* (adiabatic transformability). On the other hand, if the frequency of the oscillating field is close to the Larmor frequency, then a nonadiabatic transition is induced between states *m'≠m*. Thus the key point of this technique lies on understanding the phenomenon that takes place when the oscillating field frequency is equal to the Larmor frequency.

As mentioned earlier, Rabi *et.al.* [12] considered that a transition tooks place between two states *m* and *m'* resulting in a spatial 'reorientation' of the magnetic moment. In other words, a molecule in the homogeneous field region in a state *m* with a projection of the magnetic moment on the direction of *H₀*, *μ_z*, undergoes a transition to state *m'*. Then, it enters the inhomogeneous field region with a magnetic moment *μ_z'* (*μ_z'* ≠*μ_z*). Subsequently, the field gradient induces the molecule to change its trajectory thus it does not reach the detector.

### 2.1 New method.

There is a different mechanism for the spatial reorientation of the angular momentum that can also explain the result obtained by Rabi and Ramsey. It assumes that, in the field *C* region, there is a trajectory modification as a result of the transition between states *m* and *m'*. Therefore, the particle does not reach the detector since, in region *B*, it describes a trajectory different from the one it would have followed in state *m*.

In fact, M. Dorado [26] published a theory in which the equations of motion for systems with angular momentum were described. This theory predicts that, when the frequency of the oscillating field *H₁* is equal to the Larmor frequency, the particle feels a central force that modifies its original trajectory, which implies that the use of the inhomogeneous fields, *A* and *B*, is unnecessary.

In that work rotating coordinate systems have been used with three different reference systems.

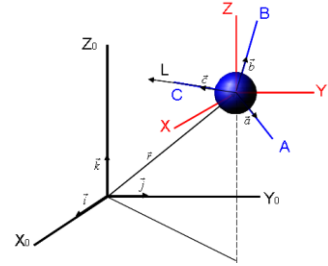

**Fig. 2.** Inertial reference system X₀ Y₀ Z₀. Non inertial reference system A B C linked to the body. Non inertial reference system X Y Z linked to the body and parallel to system X₀ Y₀ Z₀.

The equation of motion of the system in a stationary coordinate frame is

$$\hbar dL/dt = \mu \times H = \gamma_L \hbar L \times H \qquad (2)$$

In a non inertial reference system linked to the body that is rotating with angular velocity, *Ω*, around the inertial reference system, the total variation of *L* is:

$$dL/dt = \partial L/\partial t + \Omega \times L \qquad (3)$$

Where the derivative operator

$$d/dt = \partial/\partial t + \Omega \times \qquad (4)$$

has been applied. This derivative operator provides the variation of any vectorial magnitude with respect to an inertial reference system. $d/dt$ represents the evolution of the vectorial magnitude seen by a stationary observer placed in the inertial reference system. $\partial/\partial t$ represents the change of the vectorial magnitude viewed by an observer placed in the non inertial rotating system of reference frame, in relation with its own not inertial reference system. When this vectorial magnitude is the angular momentum *L* then:

$$\left(\frac{d\vec{L}}{dt}\right)_{total} = \vec{M} \qquad (5)$$

where *M* is the external torque applied over the system,

$$\left(\frac{\partial \vec{L}}{\partial t}\right) \qquad (6)$$

is the variation of *L* related to the non inertial rotating reference system and

$$\vec{\Omega} \wedge \vec{L} \qquad (7)$$

is the variation of *L* due to the rotation of the non inertial rotating reference system around the inertial reference system.

The analytical development of these equations, using these three reference systems (fig. 2), yields the general equations of motion of a rigid body which will allow us to describe its behavior.

$$M_A = \left(\frac{\partial L_A}{\partial t}\right) + \Omega_B \cdot L_C - \Omega_C \cdot L_B$$

$$M_B = \left(\frac{\partial L_B}{\partial t}\right) + \Omega_C \cdot L_A - \Omega_A \cdot L_C \qquad (8)$$

$$M_C = \left(\frac{\partial L_C}{\partial t}\right) + \Omega_A \cdot L_B - \Omega_B \cdot L_A$$

Note that, all vectorial magnitudes used here are referred to the frame linked to the body, in particular **Ω**. Besides, its components, $\Omega_A$, $\Omega_B$ and $\Omega_C$ may be written as a function of the Euler angles.

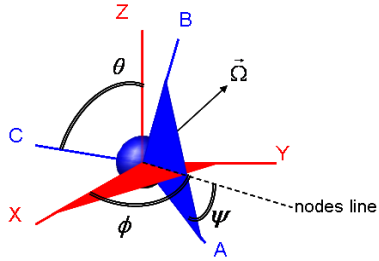

**Fig. 3.** Euler angles definiton on the systems X Y Z and ABC.

$$\Omega_A = \dot\phi \sin\theta \sin\psi + \dot\theta \cos\psi$$
$$\Omega_B = \dot\phi \sin\theta \cos\psi - \dot\theta \sin\psi \qquad (9)$$
$$\Omega_C = \dot\phi \cos\theta + \dot\psi$$

These equations (eq. 8 and eq. 9) can readily be applied to understand the effect of the rotating magnetic field used in the molecular beam magnetic resonance experiments. In our experiment there is a constant field $H_0$ around which another field, $H_1$, perpendicular to $H_0$ rotates with angular velocity **Ω**. However, from the point of view of a coordinate system rotating with $H_1$, none of the magnetic fields changes in time.

The experiment monitors, in **C** region, the interaction of the homogeneous magnetic field $H_0$ and the perpendicular oscillating magnetic field $H_1$ with the magnetic moment **μ** of a particle with angular momentum **L** as the particle moves through the fields with velocity **v**.

Therefore, the axes of the rotating coordinate frame are selected so that, at the initial instant $H_0 = H_0\,k$, $H_1 = H_1\,a$, $\mu_z = -\mu\,c$, $L_z = L\,c$, **v** = $v\,b$ and **v** can have components on *a* and *c*.

The interactions that take place at the initial instant *t* and at a later instant *t'* are described in figure 4.

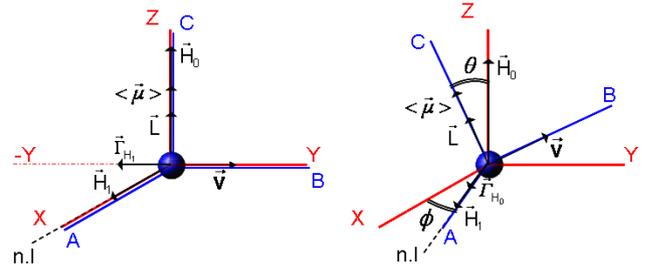

**Fig. 4.** The left figure shows the interaction of the magnetic fields $H_0$ and $H_1$ with **μ** at an instant **t**. The right figure shows the same interaction at an later instant *t'*. In both cases n.l. is the nodes line.

At the initial instant, shown on the left, the systems of axis linked to the body **ABC** and **XYZ** coincide. And for simplicity we select the **Z** axis to be parallel to $H_0$ (the homogeneous magnetic field) that defines a quantization axis. Therefore the $\mu_z$ average value and $L_z$ lay in the direction of $H_0$. In this first instant the dipolar magnetic moment and the homogeneous field do not interact since they are parallel to each other.

$$E_{H_1} = \vec H_1 \cdot \vec\mu = H_1 \cdot \mu \cdot \cos\frac{\pi}{2} = 0 \qquad (10)$$

$$\vec\Gamma_{H_1} = \vec\mu \times \vec H_1 \qquad (11)$$

However **μ** may interact with the oscillating field, $H_1$, since, **μ** and $H_1$ are perpendicular to each other. The energy of this interaction is zero but the torque $\Gamma_{H1}$ forces **L** and **μ** out of their original direction that was parallel to $H_0$.

At a later instant, *t'*, **μ** and $H_0$ can therefore interact since they are no longer parallel. The resulting energy and torque values are:

$$E_{H_0} = \vec H_0 \cdot \vec\mu = H_0 \cdot \mu \cdot \cos\theta \qquad (12)$$

$$\vec\Gamma_{H_0} = \vec\mu \times \vec H_0 \qquad (13)$$

This analysis is addressed to evaluate the evolution of the internal angular momentum **L** of the particle when a torque is applied in resonant conditions. This means that

$$\left(\frac{\partial L}{\partial t}\right) = 0 \qquad (14)$$

We only need to apply the general equations shown earlier. Using the values of the applied torques and carefully introducing the torque values in the equations $\Omega_A$ and $\Omega_B$ are obtained.

$$\Omega_A = -\frac{M_B}{L_C} = -\frac{\mu \times H_1}{L_C} \qquad (15)$$

$$\Omega_B = \frac{M_A}{L_C} = \frac{\mu \times H_0}{L_C} \qquad (16)$$

Since $M_C = 0$, means $\ddot\psi = 0$ and $\dot\psi = cte + \psi_0$. We select for convenience this constant to be cero and $\psi_0 = 0$. Using the expression of **Ω** written as function of the Euler angles, with these *contour* conditions, simplifying and substituting the values of $\Omega_A$ and $\Omega_B$ we can calculate $\dot\theta$ and $\dot\phi$ as

$$\dot\phi = \frac{\mu \cdot H_0}{L_C} \qquad (17)$$

$$\dot\theta = -\frac{\mu \cdot H_1}{L_C} \qquad (18)$$

Eq. (17) represents the Larmor frequency $\dot\phi$. In addition, our treatment yields a new component of the rotation which is the nutation frequency $\dot\theta$.

## 2.2 Rabi et al. method revisited.

A different way to solve this problem is to use two rotating coordinate systems. The extensive and explicit use of the rotating coordinate system procedures in resonance problems was introduced by Bloch, Ramsey, Rabi and Schwinger [27]. The rotating coordinate system method is equally applicable to classical and quantum mechanical systems. The equation of motion of the system in a non inertial rotating coordinate frame is

$$\hbar\left(\frac{\partial L}{\partial t}\right) = \gamma_L \hbar L(H + \dot\phi/\gamma_L) = \gamma_L \hbar L \times H_{er} \qquad (19)$$

The effective field $H_{er}$, related to the rotating coordinate system is obtained by adding the term $\dot\phi/\gamma_L$ to $H_0$.

$$H_{er} = H_0 + (\dot\phi/\gamma_L) \qquad (20)$$

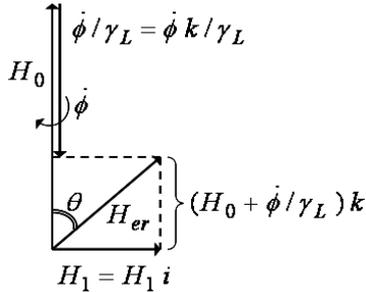

**Fig. 5.** Effective magnetic field in a rotating coordinate system. Adapted from ref.23.

The effect of a rotating magnetic field used in molecular beam magnetic or electric resonance experiments can be easily explained with this result.

In more general cases there is a constant field $H_0$ around which another field $H_1$ perpendicular to $H_0$ rotates with angular velocity $-\dot\phi$. The axes of the rotating coordinate system are selected so that, at the initial instant $H_0 = H_0\ k$, $H_1 = H_1\ a$, $\mu_z = -\mu\ c$, $L_z = L\ c$, $\dot\phi = -\dot\phi\ c$, $v = v\ b$.

Then on the rotating coordinate system

$$H_{er} = (H_0 - \dot\phi/\gamma_I)k + H_1 i \qquad (21)$$

Since the fields, $H_0$ and $H_1$, related to the rotating frame are constant in time, the solution of the motion equation of the system is much simpler in this frame than in the stationary one. The magnitude of the effective magnetic field is

$$|H|_{er} = \sqrt{\left((H_0 - \dot\phi/\gamma_L)^2 + H_1^2\right)} = |a/\gamma_L| \qquad (22)$$

Where

$$a = \sqrt{\left((\dot\phi_0 - \dot\phi)^2 + (\gamma_L H_1)^2\right)} = \sqrt{\left((\dot\phi_0 - \dot\phi)^2 + (\dot\phi_0 H_1/H_0)^2\right)} \qquad (23)$$

And, by definition,

$$\dot\phi_0 = \gamma_L H_0 \qquad (24)$$

If $\theta$ is the angle between $H_{er}$ and $H_0$ then

$$\cos\theta = (\dot\phi_o - \dot\phi)/a \qquad (25)$$

$$\sin\theta = \gamma_1 H_1/a = (\dot\phi_0 H_1/H_0)/a \qquad (26)$$

When $\dot\phi = \dot\phi_0$, $\theta = 90°$ and an electric or magnetic moment, initially parallel to $H_0$ will precess about $H_{er}$ until it becomes antiparallel to $H_0$. $\dot\phi_0$ is considered as the resonance frequency of the system. In resonance conditions:

$$a = \gamma_L H_1 \qquad (27)$$

As $\gamma_L = \mu/\hbar L$, one obtains

$$a = \mu H_1/\hbar L \qquad (28)$$

and

$$\dot\phi_0 = \mu H_0/\hbar L \qquad (29)$$

The values of $a$ and $\dot\phi$ are those obtained previously as the nutation (eq. 18) and the precession (eq. 17) frequencies respectively. This just provides us the equation of motion of the system.

## 2.3 The shaking effect.

A relevant issue is that this formalism can be applied not only to the behaviour of the angular momentum $L$, but also that of any vectorial magnitude characterizing the particle.

Following a similar method to that applied to the angular momentum (eq. 4), but in this case applied to the position vector r and knowing the value of $\Omega$, the radius of the trajectory curvature $r$ is obtained as

$$r = \frac{v}{\dot\theta} \qquad (30)$$

Once again, a similar method applied to the linear momentum and knowing the value of $\mathbf{\Omega}$ provides the momentum transfer mechanism on molecules when the resonant transition takes place.

$$\left(\frac{d(mv)}{dt}\right)_{inertial} = \left(\frac{\partial(mv)}{\partial t}\right)_{non\,inertial} + \left(\frac{\partial(mv)}{\partial t}\right)_{non\,inertial\,vs.\,inertial} \quad (31)$$

Where

$$\left(\frac{d(mv)}{dt}\right)_{inertial} = F \quad (32)$$

$$\left(\frac{\partial(mv)}{\partial t}\right)_{non\,inertial} = 0 \quad (33)$$

$$\left(\frac{\partial(mv)}{\partial t}\right)_{non\,inertial\,vs.\,inertial} = \Omega \wedge mv \quad (34)$$

and by substituting in the general equation one obtaines

$$F = \Omega \wedge mv \quad (35)$$

It is possible now to fully describe the general behaviour of the particle according to our contour and initial conditions.

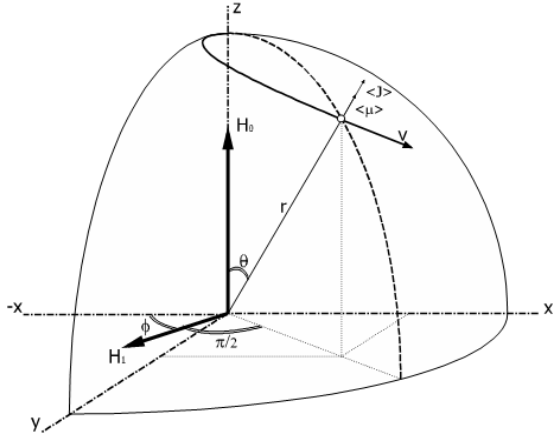

**Fig. 6.** Trajectory of the molecule when the resonant conditions are fulfilled, that is when the frequency of field $H_1$ is equal to the precession frequency (Larmor frequency) of the angular momentum $<J>$ with respect to the homogeneous field $H_0$.

There are two forces acting at the center of mass (CM) of the molecule that will modify its trajectory, from equation (35)

$$\mathbf{F}_C = m\mathbf{v}_B \times \mathbf{\Omega}_A \quad (36)$$

$$\mathbf{F}_A = m\mathbf{v}_B \times \mathbf{\Omega}_C \quad (37)$$

where $v_B$ is the velocity of the CM of the molecule. The physical interpretation is very simple. Figure 6 describes the movement of such particle when the resonant conditions are fulfilled, that is when the frequency of field $H_1$ is equal to the precession frequency (Larmor frequency) of the angular momentum with respect to the homogeneous field $H_0$.

When the interaction begins, $F_C$ induces molecules in resonance to follow a circular trajectory in the plane defined by $H_0$ and $<\mu>$, of radius $r$ (eq.30).

Simultaneously, $F_A$ causes the precession of this plane defined by $H_0$ and $<\mu>$. This effect remains as long as such molecules are under the effects of both fields, and, consequently, they will not reach the detector. For a completed development of this theory see reference [26] or [28].

Usually the radius is very small corresponding to a 'shaking' of the molecule when the transition is induced. Nevertheless it is possible to control the "shaking" process by changing $H_0$ and $H_1$.

This work suggests an experiment in which a homogeneous and a oscillating field, perpendicular to each other, act on a particle with spin and magnetic moment. That is, experimental conditions similar to those in the region *C* of the experiment carried out by Rabi. The experimental results described in this manuscript show evidences of the 'shaking' effect.

One way to determine which of the mechanisms is responsible for the observed phenomenon, either the reorientation or the trajectory modification, is to carry out a MBMR experiment and to build a experimental set up as the employed by Rabi and Ramsey but whithout using the inhomogeneous fields, *A* and *B*.

The molecular beam should be directed to the detector in a linear trajectory. If the beam deflection takes place, in resonant conditions and without the inhomogeneous fields, then the only mechanism that can explain this phenomenon is the trajectory modification. This experiment has been carried out for the hydrogen molecule and the results corroborate Dorado's theory also in agreement with previous reported works [18-24], that is, in the resonant conditions the trajectory of a molecule is modified.

## 3. Experimental Details

### 3.1 Method

The experimental method employed in this work is similar to that used by Rabi [23] in his M.B.M.R experiments, but the inhomogeneous fields have not been applied.

An experimental set up is employed where a pulsed molecular beam is generated. The pulsed valve is on a mobile system so the beam can be accurately focused onto the detector (a quadrupole spectrometer). The beam is collimated by a skimmer and on its way to the detector it passes through an homogeneous magnetic field region, $H_0$, whose direction is perpendicular to molecular beam. Additionally in this same region there is an oscillating field $H_1$ perpendicular to both the beam and $H_0$.

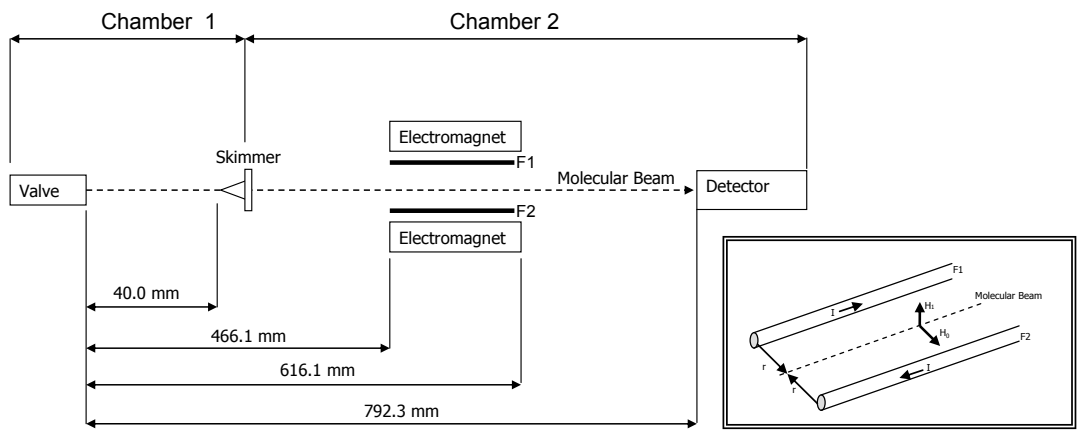

**Fig. 7**. Experimental apparatus showing the internal parts of the system divided in two high vacuum chambers i.e. pulsed valve, skimmer, electromagnet, cooper wires (F1, F2) and quadrupole detector. The inset show the configuration of the oscillating field $H_1$ and the homogeneous field $H_0$ with respect to the path of the molecular beam.

From a classic point of view, and as exposed earlier, when a molecule is in the presence of a homogeneous magnetic field, the magnetic moment of the molecule precesses with a frequency known as the Larmor frequency. From a quantum point of view, it is known that the field H quantizises the spatial position of the particle magnetic moment. Thus it restricts its orientation to certain states (Zeeman effect), so, for a transition to occur between a state m and state n, the frequency of the absorbed or emitted photon has to be

$$h \nu_{nm} = E_n - E_m \qquad (38)$$

where $E_n$ y $E_m$ are the molecule energies in states $n$ and $m$, respectively, in the presence of $H_0$. The transition from one state to the other will only take place when the $H_1$ oscillation frequency coincides with $\nu_{nm}$.

It is possible to induce the transition through two different settings. Either by choosing the oscillating frequency of $H_1$ and, varying the intensity of $H_0$, to find the energy separation in resonance with $\nu_{nm}$, Or, as in the experiments described in this work, by fixing the $H_0$ intensity and varying the oscillating frequency of $H_1$ until it coincides with $\nu_{nm}$.

### 3.2 Experimental set-up

The experimental set up, shown in Figure 7, has two ultra high vacuum (UHV) chambers. Vacuum is achieved with help of two turbo-molecular pumps (Edwards 555H) connected to either dry pre-vacuum pumps (Edwards XDS10). In the first chamber a pulsed vacuum pump injects a pressurised gas. This gas goes through the valve nozzle and it expands, generating a molecular beam. Subsequently, the beam is collimated by a skimmer (separated 40mm from the nozzle). The beam then follows a linear trajectory to reach the detector.

Inside the second chamber, and at 426.1mm from the skimmer, there is an electromagnet, which produces the homogeneous magnetic field $H_0$. This has an Armco iron nucleus annealed in a Hydrogen atmosphere. The plates are 40mm high and 150mm long and are separated 6mm. This separation remains constant throughout the experiments with a 0.05 % precision.

**Table 1.** Experimental settings.

| Pressure | |
|---|---|
| Chamber 1 | $10^{-5}$ mbar |
| Chamber 2 | $10^{-8}$ mbar |
| Pulsed Valve Nozzle | 3-4 bar |
| | |
| Molecular Beam | |
| Nozzle diameter | 0.5 mm |
| Nozzle Temperature | 78 - 310 K |
| Skimmer diameter | 0.2 mm |
| Peak velocity $v_{mp}$ | 1812 m/s |
| Beam divergence | 0.1° deg |
| | |
| Resonant unit | |
| Length | 150 mm |
| Width | 40 mm |
| $H_0$ | 1 -1800 Gauss |
| $H_1$ | 0.001- 5 Gauss |
| | |
| Radio Frequency Generator | |
| Frequency Range | 1 μHz – 20Mz |
| | |
| Quadrupole mass spectrometer | |
| Electro-ionization energy | 70 eV |
| Mass Range | 0-500 a.m.u |
| Nozzle – detector distance | 792.3 mm |
| Detector inlet diameter | 5 mm |

A current flowing through 64 hollow cooper coils generates the magnetic field. In addition, and to avoid overheating of the system, a water flow is re-circulated in the interior of the cooper wires so the magnet temperature is maintained at 25 ± 0.1 ºC with a Polyscience 6206T thermostat. To generate a weak field the electromagnet was feed with an Agilent 6611C power supply and to study the strong field range an Agilent 6684A was used.

The oscillating magnetic field $H_1$ is generated with two copper wires, 150 mm long and 1mm diameter. Both are connected in series with a hairpin shape, just as S. Millman [29] did for Rabi's original experiment. These wires are parallel to the molecular beam, so the field $H_1$ is perpendicular to $H_0$ as shown in the inset of Figure 7.

A signal generator from Agilent (33220) is used to generate the oscillating field. The pirani like detector employed by Rabi is, in this work, replaced by a quadrupole (Hidden HAL RC 511/3F), which enhances detection sensitivity and precision. In the quadrupole the incident beam is ionised by electronic impact and the registered signal is collected and displayed in an oscilloscope (Agilent Infiniium 54831B). Thus the beam velocity distribution may be analysed. The quadrupole ionization point has been set at 172,2 mm from the electromagnet exit. Table 1 summarizes the experimental conditions.

## 4. Results

The experimental results for a neutral Hydrogen molecular beam are reported here. The gas, prior to enter the vacuum chambers, is cooled with liquid Nitrogen so the rotational states more populated are the $J$=1 and $J$=0. In these conditions the Hydrogen *ortho-para* ratio is 3:1. The parallel configuration of the nuclear spin "*ortho*" is possible on even rotational states and the anti-parallel spins "*para*" on odd rotational states.

Rabi studied the resonant transitions between energetic levels of *ortho* hydrogen, $J$=1 and $I$=1, in strong field conditions, that is, in the presence of a very strong magnetic field (Back-Gouldsmit effect), where the electronic and nuclear angular momenta, $J$ and $I$ respectively, are decoupled.

In general, for the Zeeman effect, there are three experimental conditions that depend on the intensity of the homogeneous magnetic field

$\hbar\omega_0 \ll A\hbar^2$   (Weak field).
$\hbar\omega_0 \cong A\hbar^2$   (Intermediate field).
$\hbar\omega_0 \gg A\hbar^2$   (Strong field).

Where $\omega_o$ es the Larmor frequency and depends directly on the homogeneous field $H_0$ (see eq. 1), and $A$ is the coupled constant spin-orbit of the studied molecule or atom.

In particular, Rabi studied the transitions $\Delta m_I = \pm 1$ with $m_J$ = -1, 0, 1 of the Hydrogen atom for the H$_2$ molecule. In this work, these same transitions as well as the $\Delta m_J = \pm 1$ with $m_I$ = -1, 0, 1 were studied. The main difference with Rabi's work being that, in this case, the inhomogeneous fields have not been applied.

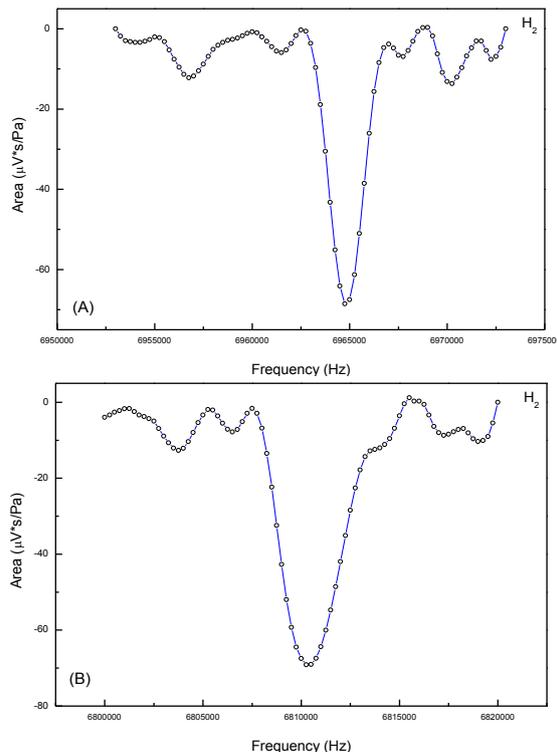

**Fig. 8.** Molecular hydrogen beam intensity measured by the quadrupole mass spectrometer set at mass 2 as function of oscillating field $H_1$. The resonant curve has been obtained in a homogeneous field $H_0$=1651 gauss for the rotational state $J$=1 and $m_J$ = -1. (A) shows the nuclear moment resonant transition from $m_I$ = -1 to $m_I$ = 0 and (B) shows the nuclear moment resonant transition from $m_I$ = 0 to $m_I$ = +1.

Figure 8 shows the resonance curves for a neutral Hydrogen molecular beam for the rotational state J=1, with $\Delta m_J = 0$ and $\Delta m_I = +1$. There, it is represented the intensity of the molecular beam that reaches the detector versus the frequency of the oscillating field $H_1$. This intensity has been determined as follows: the time of flight spectrum of the Hydrogen is integrated and normalised with respect to the working pressure in the detection chamber. The intensity of the homogeneous field $H_0$ was set to 1650.7 gauss and the intensity of oscillating field $H_1$ was set at 0.6 gauss.

It can be seen that, as the frequency of the oscillating field H$_1$ gets closer to the theoretical value of the resonance frequency between the two Zeeman levels, the intensity of the beam decreases. It is important to bear in mind that the signal intensity is directly related to a trajectory modification of the neutral molecular beam in resonance inside the homogeneous magnetic field.

The spectrum shown in section A of Figure 8 shows the transition with $m_J = -1$ $m_I'' = -1 \rightarrow m_I' = 0$ and section B depicts to the transition $m_J = -1$ $m_I'' = 0 \rightarrow m_I' = +1$.

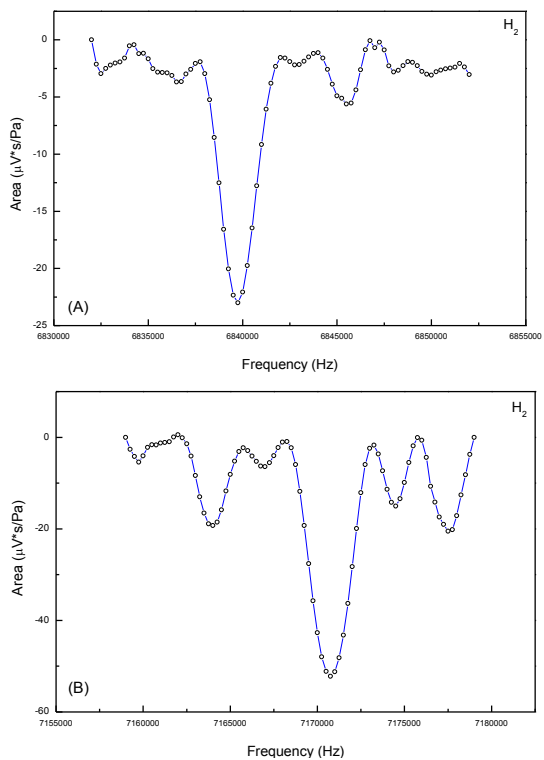

**Fig. 9.** Resonant curve for molecular hydrogen measured with $H_0$=1651 gauss for the rotational state J=1 and $m_J$ = 0. (A) shows the nuclear moment resonant transition from $m_I$ = -1 to $m_I$ = 0 and Section (B) shows the nuclear moment resonant transition from $m_I$ = 0 to $m_I$ = +1.

The interval used between two consecutive frequencies is 250 Hz. For each frequency, 30 shots from the pulsed valve were measured and the spectra shown in this work are, at least, the average of 10 spectra. Each of them including 80 different frequencies so each spectra corresponds to an average of 24000 shots of the pulsed valve.

Section A in Figure 9 shows the transition $m_J = 0$ $m_I'' = -1 \rightarrow m_I' = 0$, section *B* depicts the transition $m_J = 0$ $m_I'' = 0 \rightarrow m_I' = +1$. As in the previous figure, it can be observed that as the frequency of the oscillating field approximates that of the transition there is a decrease in the intensity.

For all the spectra an auto zeroing method was applied. This consisted of taking a reading for/a measurement for each point of the spectrum in two different conditions. Initially, the signal intensity is collected at the working frequency, and consecutively, it is read at a frequency far from the resonant frequency. The difference between these two values yields a signal intensity from which any possible interference due to the beam fluctuation have been removed.

Section A from Figure 10 represents the resonant curve for the transition $m_J = 1$ $m_I'' = -1 \rightarrow m_I' = 0$, and section *B* shows the transition $m_J = 0$ $m_I'' = 0 \rightarrow m_I' = +1$.

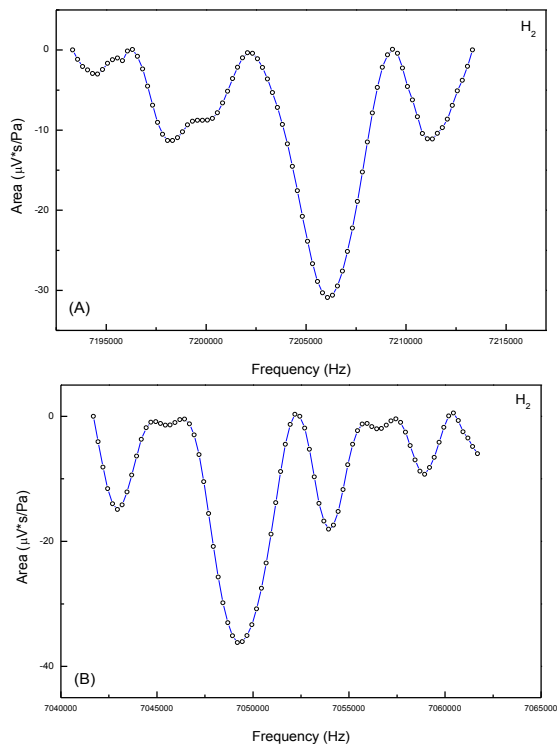

**Fig. 10.** Resonant curve for molecular hydrogen measured with $H_0$=1651 gauss in the rotational state $J$=1 and $m_J$ = +1. (A) shows the nuclear moment resonant transition from $m_I$ = -1 to $m_I$ = 0 and (B) shows the nuclear moment resonant transition from $m_I$ = 0 to $m_I$ = +1.

In adittion, as in the MBMR technique, the method described in this work has also been applied to study the series with $\Delta m_J = \pm 1$ y $\Delta m_I = 0$ for the rotational level *J*=1 and to determine the hyperfine structure on the Hydrogen molecule [30] under weak field conditions, that is to say, with coupled *I* and *J* and inducing transitions between levels with different quantum number, $M_F$. The homogeneous field varying from 100 to 400 Gauss. Results will be presented in a further publication.

## 5. Discussion

The main objective of Rabi's work [10] was to measure the nuclear magnetic moment of the proton. In fact he reported a value of 2.785 ± 0.02 nuclear magnetons. Interestingly the value currently used measured by H. S Boyne and P. A. Franken [31] is 2.79283 ± 0.00006 nuclear magnetons, in good agreement with Rabi's one.

It is interesting to compare the experimental values with the theoretical ones. The calculation of the energy of different states in ortho-hydrogen was made by Raby *et al* [10].

They assumed that the localization of these levels is similar to that of the magnetic levels in a multiplet in the Pasrchen-Back effect. Given the fact that, in this case, the spins of the two nuclei are parallel the total angular momentum (due to the nuclear spins) is 1.

$$E = -\mu_P(\sigma_1 + \sigma_2) \cdot H - \mu_R J \cdot H$$
$$- \mu_P H'(\sigma_1 + \sigma_2) \cdot J$$
$$+ \frac{\mu_P^2}{r^3}\{\sigma_1 \cdot \sigma_2 - 3(\sigma_1 \cdot r)(\sigma_2 \cdot r)/r^2\} \quad (39)$$

Where $\mu_P$ and $\mu_R$ are the proton magnetic nuclear moment and the rotational magnetic moment of the hydrogen molecule respectively (nowadays called $\mu_I$ and $\mu_J$). $\sigma_1$ and $\sigma_2$ are the spin Pauli matrices for the nuclear spins, $J$ is the rotational angular momentum, $r$ the distance between nuclei and $H$ the external magnetic field applied to the molecule. $H'$ is the field generated by the rotation of nuclei within the molecule. The results of this formalism are showed on Figure 11.

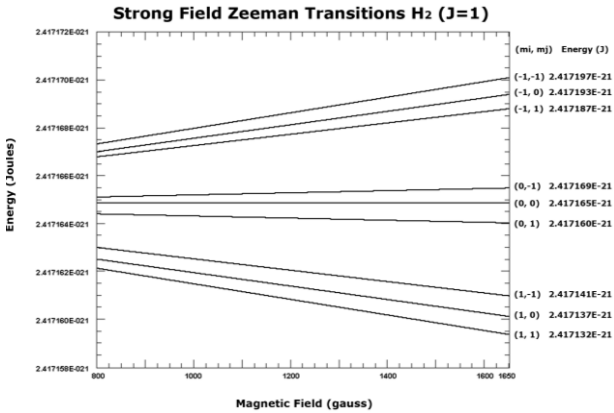

**Fig. 11.** Energy levels curve for Zeeman effect on molecular hydrogen in the rotational state J=1 under a magnetic strong field $H_0$ from 800 to 1650 gauss.

Since the interval between two consecutive points in the spectra is 250 Hz, the relative variation for the working frequencies is 0,004 %. Therefore, for those cases for which the relative difference between the theoretical and experimental values exceeds 0.004% it is possible to say that there is a significant difference between both. As indicated in Table 2, this occurs in four of the six spectra. Nevertheless, this difference is fairly small as it never exceeds 0.1%.

Using equation (1) it is possible to calculate the nuclear magnetic moment of the proton from the experimental resonant frequency. Thus here we found $\mu_P$ = 2.785 ±0.03 nuclear magnetons which is closer to the value determined by Rabi than to the one obtained by H.S. Boyne et. al [31]. However, it should be noted that Boyne's value lies within of the present result the confidence interval.

**Table 2.** Frequency values for transitions $\Delta m_I$ =0, ±1. Theoretical and experimental values.

| Transition (J = 1) | | | | $H_0$ (Gauss) | Theoretical Frequency (Hz) | Experimental Frequency (Hz) | Relative Variation |
|---|---|---|---|---|---|---|---|
| $m''_I$ | $m''_J$ | $m'_I$ | $m'_J$ | | | | |
| -1 | -1 | 0 | -1 | 1649.53 | 6 963 029 | 6 964 750 | 0,025% |
| 0 | -1 | 1 | -1 | 1650.67 | 6 810 389 | 6 810 250 | 0,002% |
| -1 | 0 | 0 | 0 | 1649.30 | 6 842 018 | 6 839 750 | 0,033% |
| 0 | 0 | 1 | 0 | 1650.53 | 7 168 778 | 7 170 750 | 0,028% |
| -1 | 1 | 0 | 1 | 1649.01 | 7 199 259 | 7 201 750 | 0,035% |
| 0 | 1 | 1 | 1 | 1650.03 | 7 046 617 | 7 046 500 | 0,002% |

Nevertheless, the main objective of the present work is to shed a light on the resonance phenomenon and attempt to explain it in terms of molecular dynamics.

**Molecule Trajectory**
M.Dorado's [26] theory predicts which would be the trajectory of a given particle with angular momentum and magnetic moment under the experimental conditions used here. That is, under the effect of a homogeneous magnetic field and a perpendicular oscillating field which is in resonance with the levels originated by the homogeneous field.

The particle movement may be described on the surface of a sphere of radius $r$ as it is plotted on Fig.6. Considering spherical coordinates, the particle will move on a meridian with angular velocity $\dot{\theta}=d\theta/dt$ and, at the same time, this meridian is precessing around the sphere with angular velocity $\dot{\phi}=d\phi/dt$.

$$r = \frac{v_0 \cdot I}{\mu \cdot H_1} \quad (40)$$

Where $v_0$ is the particle initial velocity, $I$ the angular momentum and $\mu$ the magnetic moment. $H_1$ is the oscillating field intensity. For our experimental values the trayectory radius is 2.013 m, and appling a sphere equation the variation on $z$ direction is 26 mm. The diameter of the ionization area on the quadrupole detector is 5 mm and this $z$ variation is enough to avoid it.

**Other reasons to explain the phenomenon**
Looking at the presented results it seems obvious to state that, under the resonant conditions, the intensity of the molecular beam decreases. The surprising aspect of this finding is that, in the present experiments, the inhomogeneous fields (the well known $A$ and $B$ form the MBMR) have not been employed, that is, in the experiment described here there is not a inhomogeneous field that modifies the particle trajectory.

Since the molecular beam follows a linear trajectory from the pulsed valve to the detector, a decrease in the signal intensity may be due to several reasons that should be evaluated. Those are:

*a) $H_0$ Intensity:* The homogeneous field may show some sort of inhomogeneity. If this were the case, then a fraction of the beam would not reach the detector when $H_0$ is applied, and a signal variation would be observed. However, there is no signal variation when the electromagnet is generating a 1650 Gauss field.

*b) Fringing effect on all states:* Due to the geometry of the electromagnet, similar to described on ref. [10], the gradient of the fringing effect is on the molecular beam propagation axis. This effect could increase or decrease the speed of the molecules on the progation axis direction and therefore would not explain the signal variation.

*c) Fringing effect on the final state:* Let us image a Hydrogen transition from state *m'''* to state *m''* and suppose that fringing effect only afects state *m''*. In that case the trasition from state *m''* to a new state *m'* could not be observed because molecules on state *m''* must have been previously deflected. Nevertheless transitons from state *m''* to *m'* have been shown in the results. Other reason to reject these fringing effect is that if we select a transition from *m'''* to *m''* and the $H_0$ intensity is increased then the resonant frequency is shifted but the intensity of signal depletion does not change.

*d) $H_1$ Intensity:* It could be that the oscillating field is not constant thus it generates a field gradient causing the beam to deflect. This may be discarded for two reasons; one is the low intensity of the field (0.1- 5 Gauss). The second one is that, no matter which is the intensity of this field the deflection phenomenon only occurs when the field frequency is equal to the resonant one.

*e) Deflection. vs. Depletion:* The third phenomenon to consider is beam depletion. This could be due to photodecomposition of the hydrogen molecule induced by multiphotonic absorption. However the applied energy during the experiment, ranging from 5 to 8 MHz, is far from the dissociation limit of the hydrogen molecule, $1,09265 \times 10^9$ Mhz.

*f) Photodeflection:* The transfer of the photon momentum, $h/\lambda$, associated with an absorption process, deflects the molecule of mass m with an angle $\alpha = [h / (\lambda \cdot mv)]$. For our molecular beam the result is $2.57 \cdot 10^{-12}$ radian, which is not wide enough to inhibit the Hydrogen molecules to reach the detector when the resonant transition takes place.

Therefore the only feasible possibility is that the linear trajectory of the beam is modified when the resonant phenomenon takes place. The meaning of this statement is deeper than it may look at first sight, since it implies that the reorientation mechanism proposed by Rabi and Ramsey can not be applied to explain the resonant phenomenon.

**Broadening and shape of transitions**

To study the width of the observed transitions it is necessary to consider that the experimental signals also suffer the natural line broadening due to the Doppler Effect. Other broadening sources may be the aperture time of the pulsed valve and also the one due to the electronic noise coming from the detector.

The uncertainty principle state that $\Delta E \cdot \Delta t \geq \hbar/2$, where $\Delta t$ is the time during which the perturbation is applied. Hence it is reasonably easy to deduce that $\Delta \sigma = 1/(4 \cdot \pi \cdot \Delta t)$. Fitting the velocity distribution of the molecular beam with a Maxwell-Boltzman distribution we find a velocity of 1812 m/s. The ratio between the travelled distance by the beam inside the $H_1$ region and the mean velocity yields the mean perturbation time. Therefore applying the uncertainty principle it can be said that the average width of the studied resonances is equal or greater than 961 Hz.

On the other hand, if all the transitions with $\Delta m_J = 0$ and $\Delta m_J = \pm 1$, are taken into account the average FWHM (Full With Half Maximum) of the resonant lines (listed in table 3) turns out to be 2802 Hz which completely satisfies the uncertainty principle, $\Delta \sigma \geq$ 961 Hz.

**Table 3**. Theoretical and Experimental Full With Half Maximum (FWHM) and relative intensity for transitions $\Delta m_I = 0, \pm 1$ in ortho $H_2$. This work results and Rabi's et al. [10]

| Experimental Frequency (Hz) | Experimental This Work FWHM (Hz) | Relative Intensity This Work (%) | Experimental Rabi et al. FWHM (Hz) | Relative Intensity Rabi et al. Work (%) |
|---|---|---|---|---|
| 6 964 750 | 2 011 | 18 | ~ 5 160 | 11 |
| 6 810 250 | 3 468 | 18 | ~ 5 160 | 13 |
| 6 839 750 | 2 151 | 6 | ~ 5 160 | 14 |
| 7 170 750 | 2 609 | 14 | ~ 5 160 | 12 |
| 7 201 750 | 3 408 | 8 | ~ 5 160 | 12 |
| 7 046 500 | 3 167 | 10 | ~ 4 118 | 10 |

From the energy levels calculated with equation (39), the resonant frequency for each transition is obtained. However, it should be taken into account that there is a certain probability of a resonant absorption taking place for transitions at frequencies, $\omega$, both lower and higher than $\omega_0$.

The calculation of the probability for a transition to occur in a 'gyrating' field was carried out by Güttinger [16] & Majorana [17]. They found that the probability, $P(\omega,t)$ depended, on both the magnetic field frequency and the perturbation time, *t*. Later on, I.I. Rabi [9] improved the $P(\omega,t)$ calculation, equation (41), considering that the Landé g-factor could be either positive or negative and therefore modifying the $P(\omega,t)$ value.

$$P(w,t) = \frac{(2b)^2}{(\omega-\omega_0)^2 - (2b)^2} \sin^2\left\{\frac{1}{2}\left[(\omega_0-\omega)^2 - (2b)^2\right]^{\frac{1}{2}} t\right\}$$
(41)

and

$$b = -\frac{H_1 \omega_0}{4 H_0}$$
(42)

The value of **b** is found considering an oscillating magnetic field $H_1$. If we use the approximation $\sin^2 = \frac{1}{2}$ we obtain a FWHM mean value of 2250 Hz for the six transitions ($H_0$ = 1650.0 gauss, $H_1$=0.6 gauss). A variation of ± 0.1 gauss on this aproximation produces a variation of ± 425 Hz on the FWHM. This approximation does not include our mean experimental value, since is valid only when $b \cdot t \gg \pi$ and this is far from our experimental conditions.

The theoretical FWHM must be calculated with the total probability function and considering the broadening produced when the velocity distribution of the particles in the beam is taken into account [32-33]. Hence the probability function adopts a sinusoidal form whose maximum is at $\omega = \omega_0$. The form of the resonant transitions reported here is similar to this probability function. The FWHM from equation (41) weighted with our molecular beam velocity distribution produces a mean value of 2250 Hz but in this case one variation of ± 0.1 gauss produces a FWHM of 3470 Hz for $H_1$=0.7 gauss and 1725 Hz for $H_1$=0.5 gauss. This interval includes the mean value of 2802 Hz obtained from the experimental spectra and confirms the sensibility of the line width with the intensity of oscillating field $H_1$.

The values observed by Rabi et al. [10] are different to ours (see table 3) because of two reasons: one is that their resonant cell is shorter (150 mm vs. 135 mm) and the other is that their molecular beam velocity distritution is different too.

In addition, it is also observed in all spectra that there is slight decrease in the beam intensity at both lower and higher frequencies than the resonant one.

**Intensity of line transitions**
The population of the nine energy levels for ortho-$H_2$ in the rotational state J=1 is the 75% of the total molecular beam. The depths of each minimum should be 2/9 of the total intensity and the sum of the six mimina the 12/9 of the total intensity of the beam. The expected total depletion for this six transitions should be 100% (12/9 over 75%) but on Rabi's work [10] the effect of the velocity distribution produces a 25% reduction of the intensity. This is due to variation of the transition probability per each perturbation time. In order to compare our experimental intensity with Rabi's line intensities (table 3) we normalized our results to a total value for the six transition of 75% of the total molecular beam. The statistical averaged depth for our six transition is 12.3 % ± 5.1 % which is similar to Rabi's average value, 12.4 % ± 1.4 %. The smaller standard deviation on the Rabi's messurements is explained by the use of inhomogeneus fields, that produces a reduction of the total noise and also the exclusion of 25 % para-$H_2$ contribution due to the slits position on the experimental set-up.

## 6. Conclusions

From the 30's onward, the only accepted mechanism to explain the beam deflection phenomenon observed in the MBMR is the one proposed by Rabi et al. As mentioned earlier, this explanation was based in the spatial 'reorientation' of the angular momentum when the transition between states takes place. However the 'reorientation' mechanism proposed by Rabi et al. cannot explain the results that have been obtained without the action of the inhomogeneous fields, **A** and **B**.

The new theory complete the previously explanation of the resonant phenomenun with a new mechanism to explain such phenomenon which is consistent with both the experimental results presented here and those obtained by Rabi et al.

The new method provides both, the nutation and precession frequencies. However Rabi et al. obtained only an expression for the precession frequency because they did not consider any inertial reference frame. Nevertheless comparison of both methods let us conclude that the rotating vector **a** obtained by Rabi is actually the nutation frequency.

In the new mechanism the 'reorientation' of the angular momentum is due to a change of the trajectory of the molecule determined by the nutation and the precession. This trajectory could be interpreted as a 'shaking' of the molecule.

It has been shown that the inhomogeneous fields **A** and **B** are unnecessary to induce the deflection of the resonant states of a neutral Hydrogen molecular beam. The only presence of a homogeneous and an oscillating field perpendicular to each other and perpendicular to the molecular beam is enough to induce such deflection. As a result, since the inhomogeneous fields are unnecessary to induce the deflection, the well-known MBMR technique turns out to be an easier tool to employ as the beam alignment difficulties are eliminated.

It is perhaps not surprising that, an experiment as the one described here, had never been carried out before. That is because the modification of the trajectory of a neutral molecule, in the presence of a homogeneous magnetic field, only makes sense in the framework of the new theory.

Furthermore, it is important to note that most of the experimental set ups used in MBMR studies used a beam stopper placed in the path from the oven to the detector.

Further work will include studies on the transitions of Hydrogen of hyperfine structure and will extend the study of the effect of strong fields on hetero-nuclear diatomic molecules. In fact, successful results have already been obtained for the NO molecule and will be published in a future paper.

A completed review of the theory presented in this paper has been recently published [$^{34}$].

### Acknowledgments

Authors are indebted to I.I. Rabi and N. F. Ramsey's work and respect and admire their contribution to the development of the atomic and molecular physics. We are very grateful to Prof. José Luis Sánchez Gómez from Universidad Autónoma de Madrid and Prof. Miguel Morales Furió from the Universidad Politécnica de Madrid for useful discussions. We gratefully acknowledge the contribution from E.C. Lopez-Diez, J. Andreu Serra, José Luis Pérez, and M. Sánchez Bordona from CiRTA and also from A. Montero-Dominguez, RTVE engineering deparment. We thank the Ministerio de Medio Ambiente (Project 02/2004), Parque Científico de Madrid and the Centro de Desarrollo Tecnológico Industrial (Project Neotec) for funding.